\newcommand {\di}{\partial}
\documentstyle[times,epsf,eqsecnum,preprint,aps]{revtex}
\begin{document}
\tighten
\preprint{WISC-MILW-98-TH-20}
\draft
\title{Time-Frequency Detection of Gravitational Waves} 
\author{Warren G. Anderson and R. Balasubramanian}
\address{Department of Physics, \\University of Wisconsin -- Milwaukee,\\
1900 E. Kenwood,\\ Milwaukee, WI, 53201}
\maketitle
\begin{abstract}%
We present a time-frequency method to detect gravitational wave signals in
interferometric data. This robust method can detect signals from poorly
modeled and unmodeled sources. We evaluate the method on simulated data
containing noise and signal components. The noise component approximates
initial LIGO interferometer noise. The signal components have the time and
frequency characteristics postulated by Flanagan and Hughes for binary black
hole coalescence. The signals correspond to binaries with total masses between
$45 M_\odot$ to $70 M_\odot$ and with (optimal filter) signal-to-noise ratios
of 7 to 12. The method is implementable in real time, and achieves a
coincident false alarm rate for two detectors $\approx$ 1 per 
475 years. At this false alarm rate, the single detector false dismissal rate
for our signal model is as low as 5.3\% at an SNR of 10. We expect to obtain
similar or better detection rates with this method for any signal of similar
power that satisfies certain adiabaticity criteria. Because optimal filtering
requires knowledge of the signal waveform to high precision, we argue that
this method is likely to detect signals that are undetectable by optimal
filtering, which is at present the best developed detection method for
transient sources of gravitational waves.  
\end{abstract}


\section{Introduction} \label{sec:intro}
According to Thorne\cite{300y}, gravitational wave (GW) astronomy ``will
create a revolution in our view of the universe comparable to or greater than
that which resulted from the discovery of radio waves.'' He further asserts
that ``when gravitational waves are finally seen, they will come predominantly
from sources we have not thought of or we have underestimated.'' It follows
that GW data analysis tools should include detection methods for poorly
modeled and unmodeled signal waveforms.

However, at present the only well-characterized method being widely
implemented for the detection of GWs from burst or transient sources is
Wiener's optimal filter \cite{GRASP}. This is a natural choice for sources
whose signal waveforms are theoretically well modeled, because it is the
optimal linear detection algorithm for such waveforms \cite{WainZub}.
Unfortunately, the effectiveness of optimal filtering is greatly reduced by
errors in the predicted signal waveform. Furthermore, even small errors in GW
source modeling can lead to large cumulative errors in the predicted waveform
\cite{3Min}. Optimal filtering is therefore a poor technique for inadequately
modeled (or unmodeled) sources. In fact, only two potential GW sources, binary
inspiral and black hole quasi-normal ringdown, are thought to be adequately
modeled for this method to work. Clearly, a method whose effectiveness is only
weakly dependent on (or perhaps independent of) prior knowledge of the signal
is needed. We call such methods ``robust.''

One class of robust techniques widely used for signal analysis is
time-frequency (TF) methods (cf. \cite{Boashash}). The central idea is
straightforward: one simultaneously decomposes the data in two bases, time and 
frequency. The resulting distribution, $\rho(t,f)$, represents the energy
of the data waveform at time $t$ and frequency $f$. 

Time-frequency methods are well suited to analyzing interferometric
gravitational wave data. Since interferometers are broad band instruments, the
noise energy will be distributed throughout the TF plane. GWs, on the other
hand, are caused by bulk motions of matter and energy, and their spectra tend
to be peaked about characteristic frequencies determined by the source
dynamics. GW signals may therefore be identified as ridges in the surface
$\rho(t,f)$. While preliminary investigation of time-frequency methods for GW
data\cite{GFCM,CMF,I&T1,I&T2,K&T} show promise, to date we know of no complete
TF detection method that has been implemented and evaluated.

This article describes and evaluates a time-frequency method for 
interferometric GW data analysis. The method has three steps:
\begin{enumerate}
\item transform the interferometer data into a Wigner-Ville TF distribution
$\rho(t,f)$ (described in Section \ref{sec:dist}),
\item search for ridges in $\rho(t,f)$ using Steger's algorithm (described in
Section \ref{sec:curve}),
\item threshold on length of ridge to eliminate false alarms.
\end{enumerate}
We find this algorithm can reliably detect weak signals in simulated data with
minimal assumptions about the signal.

Our evaluation of the TF method (Section \ref{sec:det}) consists of estimating
false alarm and false dismissal probabilities for a variety of signals. False
alarm probabilities are estimated by applying the method to a large number of
data segments containing only simulated initial LIGO detector noise and simply
counting detections. For false dismissal probability estimates, the same
procedure is performed on data containing both simulated initial LIGO noise
and simulated intermediate-mass ($22.5 M_\odot - 22.5 M_\odot$ to $35 M_\odot
- 35 M_\odot$) binary black hole coalescence (BBHC) waveforms. These waveforms
  are an appropriate testing ground for robust methods, since they are
probably dominated by the poorly understood merger signal. Furthermore, the
recently discovered possible evidence of ``middleweight'' black
holes\cite{Colbert&Mushotzky,Ptak&Griffiths} implies that these may be
important sources of GWs.  Flanagan and Hughes\cite{F&H} have estimated the
duration and frequency band of the merger signal, and our test waveforms are
constructed to conform to these estimates (see Appendix \ref{sec:N&S} for
details). Thus, while our model waveform is {\em not} the correct signal for a
BBHC, we believe it has the correct general characteristics for our
evaluation.

The results of our evaluation are promising. We obtain a false alarm rate of
one per 3.4 hours in a single detector, or approximately one
per 475 years for coincident detection in the two LIGO 4 km detectors. Here,
coincident means that signals are detected in both detectors within a certain
time interval, or {\em coincidence window}. The coincidence window is taken to
be 0.01 seconds (the light travel time between the LIGO 4 km detectors), since
that is the maximum time interval between the arrival of a single
gravitational wave (which travels at the speed of light) at these two
detectors. False dismissal rates vary with the signal-to-noise ratio (SNR) (as
measured by an optimal filter, cf.  \cite{GRASP}) and binary mass, but as an
example we find that approximately 3\% of signals are lost at an SNR of 11 for
a $30 M_\odot - 30 M_\odot$ BBHC.  Graphs showing false dismissal rates as a
function of binary mass for a range of SNRs and as a function of SNR for a
range of binary masses are presented in Section \ref{sec:det}. This TF method
is also computationally efficient; we are able to analyze data in these
simulations at about twice the acquisition rate of the simulated data.

While we are encouraged by these results, there is much more research to be
done. We have restricted our attention in this paper to a single TF
distribution (Wigner-Ville), a single ridge detection scheme (Steger's
algorithm), and a single thresholding scheme (ridge length). For each of these
there are a variety of choices available, and these choices need to be
explored and evaluated. We have also restricted our attention to only BBHC
signal models, and only for a limited binary mass range. While we feel that
these are priority targets for such a robust search method, there are other
sources which should be investigated. A more complete description of these and
other issues for future research are presented in Section \ref{sec:conc}.

\section{The Wigner-Ville Distribution} \label{sec:dist}
\subsection{Time-Frequency Distributions}
The central idea of TF methods is to convert time domain data, $h(t)$, into a
time-frequency distribution (TFD) $\rho(t,f)$. Here, the variable $f$ labels
the basis vectors in an alternative basis which spans the Hilbert space to
which $h(t)$ belongs (i.e. alternative to a basis labeled by $t$). This
alternative basis is usually the frequency basis associated with the Fourier
representation of $h(t)$, although it need not be (e.g. when wavelet
bases are used). Likewise, the meaning of $\rho(t,f)$ varies, although it is
usually associated in some way with the squared magnitude of $h$ at time $t$
and frequency $f$.

Methods for constructing $\rho(t,f)$ can be divided into two broad
categories\cite{Flandrin}: {\em atomic decompositions} such as windowed
Fourier transforms and wavelet transforms and {\em bilinear distributions} of
the form
\begin{equation}
   \rho(t,f)=\int\int h(u)~ h^{*}(v)~W(u,v;t,f)~du~dv
\label{eq:bidist}
\end{equation} 
where $*$ denotes complex conjugation and the kernel $W$ satisfies
\begin{equation}
   \int\int W(u,v;t,f)~dt~df = \delta(u-v).
\label{eq:kerncond}
\end{equation}
Both types of TFD might in principle be used to detect gravitational waves.

The choice of a suitable TFD is governed by two requirements. The first is
that the TFD must have good time-frequency localization properties. By this we
mean the following: if a signal written in the form $h(t)=A(t)\cos\varphi(t)$
(with $\varphi(t)$ a continuously differentiable monotonic function) satisfies
the adiabaticity conditions
\begin{eqnarray}
   A\left(\rule{0cm}{4mm}t+\delta(t)\right)\approx A(t), 
\\ 
   \dot{\varphi}\left(\rule{0cm}{4mm}t+\delta(t)\right)\approx 
      \dot{\varphi}(t),
\label{eq:locconds} 
\end{eqnarray} 
where $\delta(t)$ is defined by
$\varphi\left(\rule{0cm}{4mm}t+\delta(t)\right)=\varphi(t)+2\pi$ for all $t$,
then it should give rise to a distribution $\rho(t,f)$ which has support only
in a small neighborhood of the curve $f=\dot{\varphi}(t)$. In other words, if
the signal has a well defined frequency at each time, then the TFD should
reflect that. This is required for a ridge detection algorithm to work well
with the TFD. The second requirement is that the TFD be relatively inexpensive
to compute. This makes real-time analysis of interferometer data feasible.

Gon\c{c}alves, Flandrin and Chassande-Mottin\cite{GFCM} have investigated the
localization properties of various TFDs when applied to binary inspiral chirp
waveforms. They conclude that bilinear distributions have superior
localization properties, and we therefore focus our attention on those. On the
other hand, bilinear distributions can be computationally expensive, depending
on the choice of kernel $W$ in (\ref{eq:bidist}). The question, then, is
whether there is a kernel for which $\rho$ can be calculated efficiently.

\subsection{The Wigner-Ville Distribution}
There is at least one choice of kernel which leads to a computationally
efficient bilinear distribution with good localization properties: the {\em
Wigner-Ville distribution} (WVD)\cite{Wigner,Ville}
\begin{equation}
   \rho(t,f)=\int_{-\infty}^{\infty} h^*\left(t-\frac{\tau}{2}\right)
      h\left(t+\frac{\tau}{2}\right)e^{i2\pi f \tau} d\tau.
\label{eq:WigTrans}
\end{equation} 
We will use the WVD , but note that it exhibits two features that might at
first glance seem undesirable.

Both features are illustrated by considering a purely
sinusoidal signal, $h(t)=\sin(2\pi f_0 t)$. The WVD for this $h(t)$ is
\begin{equation} 
   \rho(t,f)=\frac{1}{4}\left[\delta(f-f_0)+\delta(f+f_0)-
      2\cos(4 \pi f_0 t)\delta(f)\right].  
\label{eq:WigSin} 
\end{equation} 
First, note that $\rho(t,f)$ does not have support only at the sinusoid's
frequency $f=\pm f_0$, but also has an ``echo'' at $f=0$. The sum of two
sinusoids $h(t)=\sin(2\pi f_1 t) + \sin(2\pi f_2 t)$, has further echoes at
$f=\pm(f_1\pm f_2)$. These echos are due to the bilinearity of the
distribution. Second, observe that this $\rho(t,f)$ attains negative values at
$f=0$ whenever $\cos(4 \pi f_0 t)$ is positive.  Thus, it is not entirely
correct to think of the $\rho(t,f)$ as the squared magnitude of the signal at
a give time and frequency. 

Despite this behavior, the WVD is well suited to our method. We wish to
look for ridges in the TF plane. Negative values away from these ridges are of
no consequence. In fact, throughout this paper we will adopt the
convention of setting negative values of $\rho(t,f)$ to $0$. Also, since we
are interested only in detection here, and not in extracting information about
the signal, it does not matter how many ridges there are in the TFD, it
matters only that their existence be highly correlated with the presence of a
signal and that they be sharp. The WVD satisfies both these criteria. 

\subsection{Discrete Implementation}
While the discussion above has been couched in the language of functions of
continuous variables, in practice it is necessary to calculate WVDs from
discretely sampled data $h_j=h(j\Delta t)$. This appears to presents a minor
dilemma, since (\ref{eq:WigTrans}) contains expressions of the form
$h(t-\tau/2)$, which when discretized become $h_{j-k/2}$. For interferometer
data, this issue is easily dealt with, since the data will be significantly
($\sim$ 10 times) oversampled. It is thus a simple matter to decimate these
data to twice the time resolution required. We resample our simulated data so
that $h_j\equiv h(2j\Delta t)$ accordingly.

A second issue (also present in the continuous case) is that in practice
$h(t)$ is known only in a time interval $0\le t \le T$.  This means that
$h(t-\tau/2)$ is undefined for $\tau/2>t$, and likewise $h(t+\tau/2)$
for $\tau/2>T-t$. In order to calculate (\ref{eq:WigTrans}) one must assign
values to $h(t)$ in the entire time interval $-\infty \le t \le \infty$. 
We therefore take $h(t)=0$ in the intervals $t<0$ and $t>T$.

Having resolved these issues, it is straightforward to calculate the discrete
analog of (\ref{eq:WigTrans})
\begin{equation}
   \rho_{jk} = \sum_{\ell=-N/2}^{N/2}~h_{(j-\ell/2)}~h_{(j+\ell/2)}
      ~e^{2\pi i k \ell/N},
\label{eq:DWVD}
\end{equation}
where $N=T/\Delta t$. This discrete distribution can be treated as a digital
image, with $j$ and $k$ respectively denoting the horizontal and vertical
pixel number. The value of $\rho_{jk}$ denotes the gray scale value, or 
equivalently height (i.e. $z$ value), of the image in that pixel. 

It is easy to estimate the computational efficiency of the discrete WVD from
(\ref{eq:DWVD}). For each value of $j$, one does a single multiplication (of
negligible cost) and a Fourier transform. The Fast Fourier transform costs 
$N\log_2(N)$ floating point operations, and must be done for each of $N/2$
possible values of $j$. Thus, the cost of calculating the WVD is approximately
$(N^2 \log_2 N)/2$.

Finally, note that for a data set with $N$ samples the resulting WVD has
$N^2/8$ pixels ($N/2$ time intervals by $N/4$ positive frequency bins).
Images from large data sets therefore quickly become unwieldy; for example the
data sets we used contained 4096 floats, leading to an image of $2048
\times 1024$ pixels. We therefore averaged over 4 pixel intervals in time and
2 pixel intervals in frequency to obtain a more wieldy image size of $512
\times 512$. 

\begin{figure}[h]
\epsfxsize15cm
\epsffile{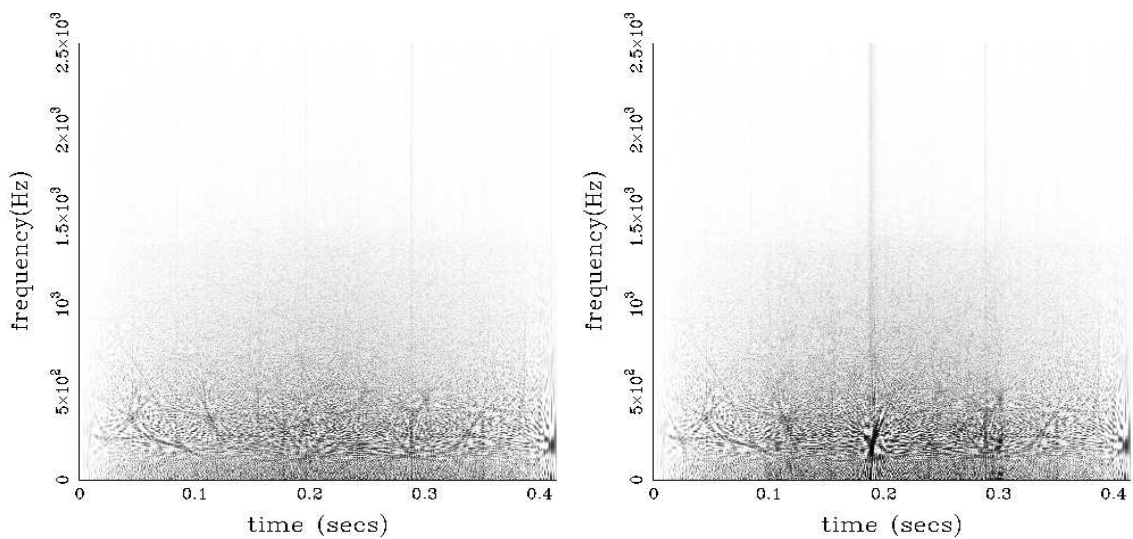}
\caption{The WVDs of a) simulated initial LIGO noise and b) a simulated
signal and noise (optimal filter SNR=8). In both cases, the data is
over-whitened (see Section \ref{sec:det}). In this figure, the darkness of a
pixel corresponds to the magnitude of $\rho(t,f)$ in that pixel, with darker
pixels having higher values. Thus, a dark curve corresponds to a ridge in the
distribution. In b), the signal, which is a model binary black hole
coalescence waveform, is the dark structure in the bottom center (at about 0.2
sec) of the figure. The additional structure in the figures is self
interference due to the bilinearity of the Wigner-Ville distribution.
\label{fig:WigEx}}
\end{figure}

We end this Section with an example of a discrete Wigner-Ville TFD.  Figure
\ref{fig:WigEx} shows two WVDs: one of simulated initial LIGO noise and the
other of a simulated BBHC waveform embedded in that same noise
(see Section \ref{subsec:Noise} and Appendix \ref{sec:N&S} for details).

\section{Steger's Ridge Detection Algorithm} \label{sec:curve}
A gravitational wave signal in interferometer data $h(t)$ should produce a
ridge in the TFD $\rho(t,f)$.  Therefore, to detect GWs, a ridge detection
algorithm (or equivalently line detection algorithm if $\rho(t,f)$ is
represented as a gray-scale map as in Fig.  \ref{fig:WigEx}) is required.
Fortunately, there are a number of ridge detection algorithms in the digital
image processing and computer vision literature from which to choose
\cite{H&S}.  

We use Steger's second-derivative hysteresis-threshold algorithm
\cite{steger98}. The essential idea of this scheme is simple. A ridge in a
surface will have high curvature (second derivative of $\rho(t,f)$) in the
direction perpendicular to the ridge. Furthermore, the first derivative will
vanish at the top of the ridge, since it is a local maximum. Thus, ridges are
identified as contiguous sets of point at which $\rho(t,f)$ has a
high-curvature local maximum.

Steger's algorithm, which identifies ridges in this way, has a number of
steps. First the TFD $\rho(t,f)$ is convolved with a 2-dimensional Gaussian
smoothing kernel,
\begin{equation}
   K(t,f)=\frac{1}{2\pi\sigma^2}~e^{-\frac{1}{2}\left[\left(
      \frac{t}{\sigma t_p}\right)^2 +\left(\frac{f}{\sigma f_p}\right)^2
      \right]},
\label{eq:StegKern}
\end{equation}
where the dimensionless scale parameter $\sigma$ allows the preferential
detection of ridges of width \raisebox{-.8ex}{$\stackrel{<}{\sim}$} $\sigma
t_p$ or$\sigma f_p$ , and the parameters $t_p$ and $f_p$ are characteristic
time and frequency scales of the TFD (we used $t_p = 4 \Delta t$ and $f_p=
2/T$, the time and frequency resolution of the digitized $\rho(t,f)$). One
then takes the the first and second derivatives of the convolution
\begin{equation}
   H(t,f)=\int^\infty_{-\infty}\int^\infty_{-\infty}\rho(t',f')K(t-t',f-f')
      ~dt'df',
\label{eq:KernConv}
\end{equation} 
with respect to both $t$ and $f$. The second derivatives are used to find the
eigenvectors $(n_t,n_f)$ of the Hessian matrix,
\begin{equation}
   \left(\begin{array}{cc}
      \frac{\di^2}{\di t^2}&\frac{\di^2}{\di t \di f}\\
      \frac{\di^2 }{\di f \di t}&\frac{\di^2}{\di f^2}
   \end{array} \right)H(t,f) 
   \left(\begin{array}{c}
      n_t\\
      n_f
   \end{array} \right) = \alpha
   \left(\begin{array}{c}
      n_t\\
      n_f
   \end{array} \right).
\label{eq:Hessian}
\end{equation}
At each point in the $t-f$ plane, the eigenvector corresponding to the largest
eigenvalue magnitude, $|\alpha|$, defines the line in the $t-f$ plane along
which the second derivative of $H(t,f)$ obtains its extremal value. For points
on a ridge, these vectors will be normal to the curve described by the ridge
in the $t-f$ plane. If the first derivative of $H(t,f)$ vanishes in this
direction,
\begin{equation}
   \left(n_t \frac{\di}{\di t} + n_f \frac{\di}{\di f} \right) H(t,f) =0,
\label{eq:linecond}
\end{equation}
then the point may be on a ridge. We call such points ``potential ridge 
points''. 

If all potential ridge points were included in ridges, one would find ridges
everywhere in the TFD due to noise fluctuations, whether a signal was present
or not. To avoid this, one thresholds on the value of the second derivative.
However, thresholding presents conflicting requirements. On one hand, if the
threshold is too low, it will allow too many noise ridges to be detected. On
the other hand, if it is too high, portions of a genuine signal ridge may be
missed due to deviations below the threshold caused by noise. To improve the
detection of signal ridges while suppressing noise ridges, hysteresis
thresholding is used. This means that there are two thresholds on the second
derivative of $\rho$, an upper threshold, which must be exceeded by any point
at which a ridge can start, and a lower threshold, which must be exceeded by
every point on the ridge. 

Finally, ridges are identified as contiguous sets of potential ridge points
which meet the hysteresis thresholding criteria. Isolated potential ridge
points are not defined to be ridges. Note that small gaps in a ridge will be
``smeared out'' when the TFD is convolved with the Gaussian kernel. Thus, gaps
of less than a few $\sigma f_p$ or $\sigma t_p$ are overlooked by this
algorithm, which decreases the probability of noise fluctuations breaking a
signal ridge into many smaller ridges. Conversely, the minimum ridge length
detected is also a few $\sigma t_p$ or $\sigma f_p$, eliminating many 
shorter noise ridges.

There are some further issues involved with implementing this algorithm on a
digital image. When $t$ and $f$ are replaced by their discrete counterparts,
$t_j$ and $f_k$, the distribution $\rho_{jk}=\rho(t_j,f_k)$ can be viewed as a
piecewise constant function, having value $\rho_{jk}$ in the rectangle
$[t_j-(\Delta t/2),t_j+(\Delta t/2)] \times [f_k-(\Delta f/2),f_k+ (\Delta
f/2)]$, where $\Delta t = t_j - t_{j-1}$ and $\Delta f = f_k - f_{k-1}$, and
vanishing outside the image (i.e. when $j<0$ or $k<0$ and when $j>N$ or
$k>N$).  This piecewise constant function is then convolved with the {\em
continuous} Gaussian kernel. The convolution is equivalent to the summation
\begin{equation}
   H_{mn}=\sum_{j} \sum_{k} \rho_{jk} K_{(j-m)(k-n)},
\label{eq:DiscConv}
\end{equation} 
where the convolution mask $K_{(j-m)(k-n)}$ is given by
\begin{equation}
   K_{(j-m)(k-n)}=\int_{t_j-t_m-\Delta t}^{t_j-t_m+\Delta t}
      \int_{f_k-f_n-\Delta f}^{f_k-f_n+\Delta f}
      K(t,f)~dt~df.
\label{eq:DiscMask}
\end{equation}
Rather than taking discrete derivatives of $H$, the process can be made more
efficient by using masks built from the derivatives of $K$. By integrating
(\ref{eq:KernConv}) by parts, one may see that the same result can be derived
by either process. However, while the derivatives of $H$ need to be calculated
at each step, the derivatives of $K$ can be calculated just once.  Thus, along
with (\ref{eq:DiscMask}), convolution masks are also made with $K$ replaced by
its first and second derivatives with respect to $t$ and $f$.

Another issue addressed by Steger's algorithm is the positioning of potential
ridge points. For a digital image, the ridge will be composed of pixels within
which the directional derivative (\ref{eq:linecond}) vanishes.  A one
dimensional example illustrates the method used to determine whether
(\ref{eq:linecond}) is satisfied within a given pixel. Denote the
one-dimensional image function by $f(x)$. Approximate $f(x)$ at a point $x_j$
by its second order Taylor series,
\begin{equation}
   f(x)\approx f_j + \frac{1}{\Delta x}(f_{j+1}-f_{j-1})~(x-x_j)+ 
      \frac{1}{2(\Delta x)^2} (f_{j-1}+f_{j+1}-2f_j)~(x-x_j)^2,
\label{eq:Taylorf}
\end{equation}
where the coefficients are finite difference approximations of the derivatives
of $f$ at $x_j$ and $\Delta x = x_{j+1}-x_j$. The derivative of
(\ref{eq:Taylorf}) vanishes at $x=x_j+dx$ where
\begin{equation}
   dx\equiv-\Delta x \left(\frac{f_{j+1}-f_{j-1}}{f_{j-1}+f_{j+1}-2f_j}\right).
\label{eq:DiscLinePt}
\end{equation}
This is within the $j^{th}$ pixel if and only if $x_j-\Delta x/2 < x_j+dx < 
x_j+\Delta x/2$. In that case, $x_j+dx$ is considered to be a potential ridge 
point and the $j^{th}$ pixel is a ``potential ridge pixel.'' One then joins
ridge pixels (as before) to find the ridge. The generalization to two
dimensions is discussed in \cite{steger98}.

\section{The Detection Method}\label{sec:method}
We have thus far described two of the three steps in our GW detection method:
generation of WVDs from interferometric data and the search for ridges
(signals) in them. If all ridges corresponded to GW signals, the existence of
a ridge would be an adequate detection statistic and these two steps would be
sufficient. However, even after hysteresis thresholding (see Section
\ref{sec:curve}), stochastic detector noise will lead to spurious ridges in
the WVD. This may be exacerbated if the noise is non-stationary and/or
non-Gaussian, in which case signal ridges and noise ridges might have very
similar characteristics.  A more powerful statistic can be devised by
thresholding on ridge characteristics which are more strongly correlated with
signal ridges than with noise ridges.  Of course, the more specific the
threshold is to the signal, the less robust the detection technique will be. 

Ridge length (i.e. the number of pixels in a ridge) is one characteristic
which distinguishes signal ridges from noise ridges in stationary Gaussian
noise. Since the noise is stochastic, it will produce a distribution of ridge
lengths, with short ridges being more frequent than longer ones (e.g. see Fig.
\ref{fig:FA}). The ridge length of a signal is (approximately) determined by
the signal's duration and frequency band.  Thus ridge length is more strongly
correlated with signal ridges than noise ridges, and setting a minimum ridge
length threshold can further improve the detection statistic. Also, note that
this thresholding is not so specific as to undermine the robustness of the TF
method. For these reasons, we use ridge length as a threshold. 

Our complete TF detection method consists of the following steps:
\begin{enumerate}
\item construct the Wigner-Ville distribution $\rho(t,f)$ of the detector 
output $h(t)$ as per Section \ref{sec:dist},
\item search for ridges in the TF map $\rho(t,f)$ using the Steger's algorithm
as per Section \ref{sec:curve} 
\item use a threshold on the length of the ridge as a detection criterion.
\end{enumerate}
The remainder of this article evaluates the performance of this method.

\section{Signal Detection and False Alarm Statistics} \label{sec:det}
The statistical performance of our TF method is measured by two probabilities: 
the probability of finding a signal when there is none, or {\em false alarm 
probability} and the probability of failing to find a signal when there is
one, or {\em false dismissal probability}. These probabilities depend on the 
details of the noise and the signal. 
\subsection{Noise and Signal Models} \label{subsec:Noise}
Our simulations were carried out with discretely sampled colored
Gaussian noise. This noise was produced using the {\em Numerical Recipes}
\cite{numrep} routine {\tt gasdev$\!$()}. We altered {\tt gasdev$\!$()} to
use the {\tt ran2$\!$()} random number generator, since it produces much
longer sequences of pseudo-random numbers than the default {\tt ran1$\!$()}
routine. We divided the unit-variance white Gaussian deviates produced by {\tt
gasdev$\!$()} into two equal sets, which were used as the real and imaginary
components of the complex valued frequency domain noise data, $\tilde{n}_j$.
Finally, we colored the noise data by multiplying them by the square root of
the initial LIGO noise curve, $\tilde{n}_j\sqrt{S_h(|f|)} \rightarrow
\tilde{n}_j$. The specific $S_h(|f|)$ we use is that of Abramovici et al.
\cite{Abramovicietal} as generated by the GRASP routine {\tt
noise\_power$\!$()} with input parameter {\tt
noise\_file="noise\_ligo\_init.dat"} \cite{GRASP}.

\begin{figure}
\epsffile{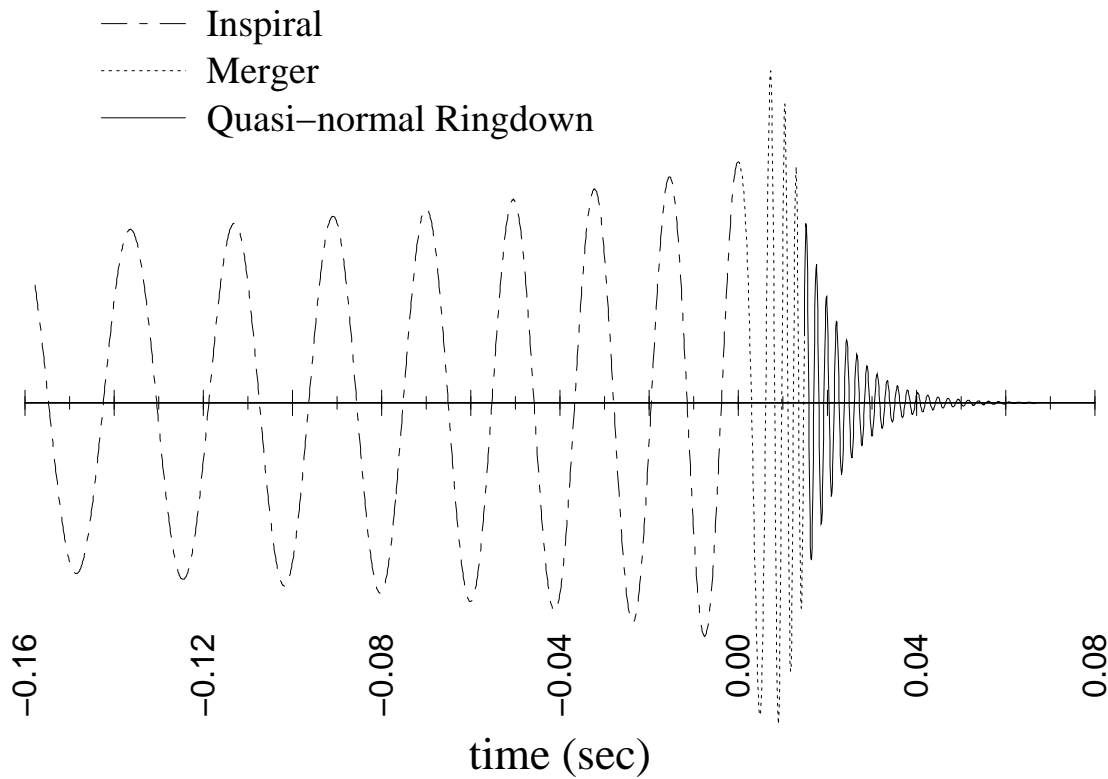}
\caption{A simulated binary black hole coalescence waveform for a 
$30 M_\odot$ - $30 M_\odot$ system, showing the three phases of the
coalescence described by Flanagan and Hughes \protect\cite{F&H}\protect. The
frequency of the wave increases monotonically. Only the portion of the
waveform with frequency $\ge$ 40 Hz is shown.
\label{fig:BBHwave}}
\end{figure}

The signal data was obtained by sampling a continuous waveform developed
specifically to test this algorithm. This waveform simulates the signal from
an intermediate-mass coalescing black hole binary. These are expected to be
important sources whose gravitational waveforms cannot be calculated
analytically in the initial LIGO sensitivity band. They are thus ideal
candidates for a robust detection method. Our model is based on the
predictions of Flanagan and Hughes \cite{F&H}. We wish to emphasize that it is
{\em not} intended to be ``accurate'' in the sense required for optimal
filtering: rather, it is constructed to have time, frequency and energy
characteristics consistent with the assumptions of \cite{F&H}. Thus, the
performance of our TF method for these simulated signals should be indicative
of its performance for actual coalescence signals. A detailed description of
the model waveform and its derivation is presented in Appendix \ref{sec:N&S}.
A typical waveform (for a $30 M_\odot - 30 M_\odot$ binary system) is shown in
Fig.~\ref{fig:BBHwave}.
\subsection{The Simulation Procedure}
We analyzed simulated data in segments of $N = 4096$ samples each. The assumed
sampling frequency was $f_s = 9868.420898$Hz (we chose the sampling frequency
of the LIGO 40-meter prototype detector so that data from that detector could
be easily analyzed). Each segment therefore lasts for $T = N/f_s \approx
0.415$ s, and the time interval between successive samples is $\Delta t =
1/f_s \approx 1.013\times10^{-4}$ s. Samples in each segment are denoted by
$h_j = h(j\Delta t)$ where $j$ takes the values between $0$ and $4095$. We
denote the Fourier transform of $h(t)$ by $\tilde h(f)$ and its discrete
representation by $\tilde h_j = \tilde h(j\Delta f)$, where $\Delta f = 1/T$
and $j$ takes the value between $0$ (DC) and $N/2$ (the Nyquist frequency). 

We performed two types of simulations: one to determine false alarms
probabilities (the probability that a signal is detected when none is present)
and one to determine false dismissal probabilities (the probability that no
signal is detected when one is present). In every simulation, the data stream
contained Gaussian noise, as described in Subsection \ref{subsec:Noise}. To
determine false alarm probabilities, no simulated signals were added, so that
$\tilde{h}_j=\tilde{n}_j$. To determine false dismissal probabilities,  
simulated GW signals were added to the noise. 

As indicated in Appendix \ref{sec:N&S}, the angle averaged signals were
generated in the time domain. We therefore took the Fourier transform of the
average signal, $\langle \tilde h^{\rm BBH} \rangle_k \equiv
\sum_{j=0}^{N-1}e^{2\pi ijk/N}\langle h^{\rm BBH} \rangle_j $. The signal was
normalized so that
\begin{equation}
\sum_{k=0}^{N/2}\frac{\left|\langle \tilde h_{\rm BBH}
\rangle_k\right|^2}{(S_h)_k} = 1,
\end{equation}
where $(S_h)_k=S_h(k\Delta f)$ is the one-sided power spectrum of the noise. 
The frequency domain data stream is then taken to be,
\begin{equation}
\tilde h_k = \tilde n_k + {\rm SNR}\, \langle \tilde h^{\rm BBH}\rangle_k,
\end{equation}
where SNR is the signal-to-noise ratio obtainable by matched filtering.

Since the Wigner-Ville distribution is computed from the time series
representation of the detector output, it was necessary to transform $\tilde
h_k$ into the time domain. However, the power spectrum spans such a large
dynamic range that significant numerical errors arise if one simply takes the
inverse Fourier transform of $\tilde h_k$. We therefore ``over-whitened'' the
data, $\tilde h_k, \rightarrow \tilde h_k / (S_h)_k,$ to reduce the
(time-domain) dynamic range. This procedure has the added benefit that it
emphasizes the frequencies in which the detector is most sensitive while
suppressing the frequencies where the noise is large. We then took the inverse
Fourier transform of the over-whitened $\tilde h_k$, and subsequently computed
its WVD, $\rho_{jk}\equiv \rho(j\Delta t,k\Delta f)$, as per Eq.
(\ref{eq:WigTrans}). 

This procedure produces discrete TFDs, whose dimensions may be expressed in
units of pixels. Because of the undersampling involved in computing the 
discrete WVD (see Section \ref{sec:dist}) we computed
$\rho_{jk}$ at $4096/2=2048$ distinct offsets, giving the WVD a ``width'' of
2048 time pixels. Also, the transform is constructed only at positive
frequencies up to the Nyquist frequency {\em after} undersampling. Thus, the
WVD has a ``height'' of 1024 frequency pixels. The resulting TFDs had
$2048\times1024\approx2\times10^6$ pixels. This was too large for efficient
computation, so we reduced the map size by averaging over every 4 time pixels
and every 2 frequency pixels. The final discrete WVDs had dimension
$512\times512$ pixels, each pixel having area $0.0008 s \times 9.6 Hz =
7.8 \times 10^{-3}$.

Next, $\rho_{jk}$ was passed through Steger's line recognition algorithm. We
used Steger's implementation of this algorithm\cite{StegCode}, which assumes
that the pixels of the image (distribution in our case) are unsigned
characters, taking integer values between 0 and 255. To convert the floating
point image $\rho_{jk}$ to its unsigned character analog ${(\rho^{b})}_{jk}$
it was necessary to rescale the data to fit in this range without saturating it.
The scaling factor was calculated as follows: the maximum floating point value
of $\rho_{jk}$ was found for each segment of noise data. The scaling factor
was then chosen so that it scaled the average value of these maxima to the
value 128. Once rescaled, the image was passed to Steger's algorithm. The
algorithm parameters we used were $\sigma=2$ and second derivative
(hysteresis) thresholds of $10/{\rm pixel}^2$ and $3.33/ {\rm pixel}^2$.
Ideally, these values would be chosen by some optimization procedure, however,
for this preliminary study they were selected ``by hand'' after extensive
numerical experimentation. 

For each map, the line recognition algorithm returns a list of ridges detected
and their lengths. Our detection statistic was the length of the longest ridge
in the map. Thus, a threshold value was chosen, and if this value was equaled
or exceeded by the longest curve in a given map, a signal was said to have
been detected in that map. For instance, if one chooses a threshold ridge
length 30 pixels, a signal is said to have been detected in any map which
contains a ridge whose length is 30 pixels or longer.
\subsection{False Alarm Probability}
Our goal is an algorithm which has an acceptable false alarm rate $R_f$. In
our case, this means determining a ridge length which is not equaled or
exceeded in maps containing only noise more than once in every $1/R_f$
seconds. To compute this threshold, we simulated noise from an ensemble of
statistically independent identical detectors. In our simulation, the ensemble
consisted of approximately $1.7\times 10^6$ detectors, which corresponds to
analyzing $1.7\times10^6~\times~0.415$ seconds or about 196 hours of simulated
data. 

\begin{figure}
\epsffile{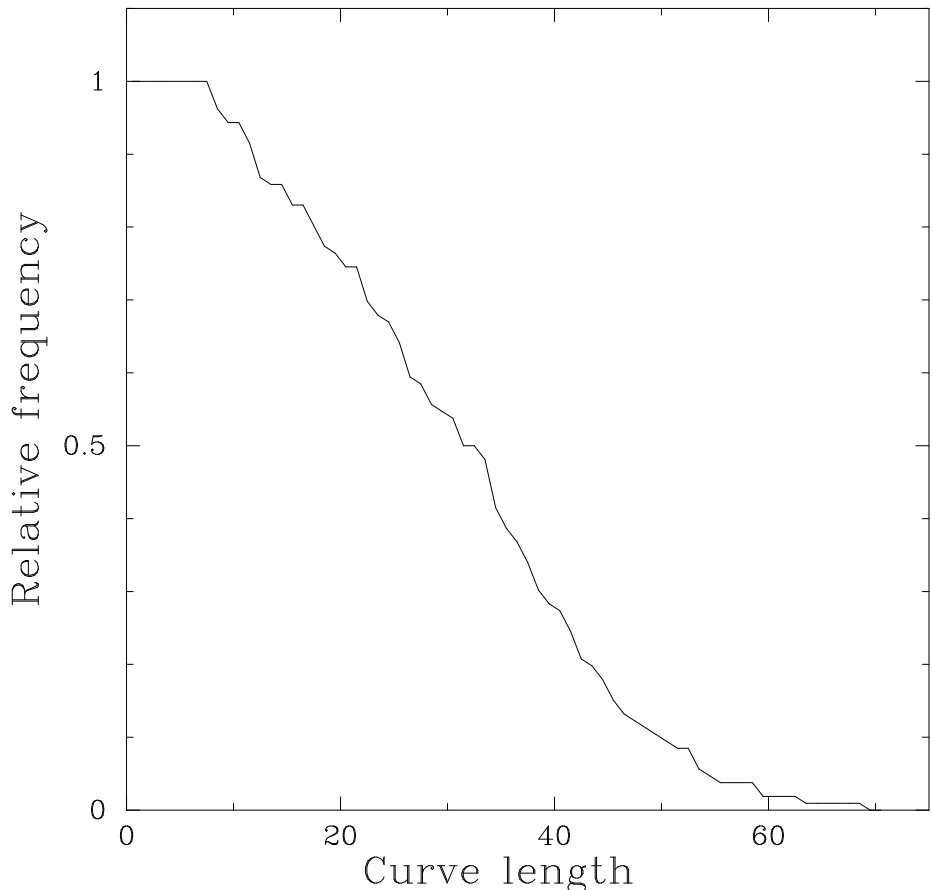}
\caption{The relative frequency with which Steger's algorithm detected ridges
with length $\ge \ell$ pixels in WVDs of simulated initial LIGO noise. Only
the 106 maps in which ridges were detected are included in this graph. The
unevenness of the graph is caused by statistical fluctuations due to small
number statistics.
\label{fig:FA}}
\end{figure}

For each member of the ensemble, we computed a WVD. We then searched for
ridges in the WVD using Steger's algorithm and determined the length of the
longest ridge. Ridges were found in 106 of the $1.7\times 10^6$ WVDs, or
about 1 out of every $16\,000$ maps. In those maps in which ridges were found,
the length of the longest ridge ranged from 7 to 68 pixels. The relative
frequency with which the longest ridge had a length $\ge \ell$ pixels is shown
as a function of $\ell$ in Figure \ref{fig:FA}. 

Based on this simulation we chose a ridge length threshold of 30 pixels. This
threshold produces a false alarm probability of about $3.4\times10^{-5}$, or a
false alarm rate of about one every $3.4$ hours in a single detector. More
interesting is the rate at which false alarms occur simultaneously in two
uncorrelated detectors. This coincidence rate is proportional to the
coincidence window $\tau$, i.e. the time interval within which the same signal
can be seen by both detectors. A natural lower limit for $\tau$ is the light
crossing time between detectors, $\tau_c$, since the time interval between the
arrival of a single signal at the first detector and the last detector can be
up to $\tau_c$. However, other considerations may set a larger $\tau$ for a
given algorithm, leading to a higher coincident false alarm rate.

For the LIGO 4 km detectors, $\tau_c$ is approximately 10 ms. As implemented,
the TF method did not resolve time-of-arrival to this accuracy, because we did
not distinguish when in the 400 ms data segment the signal arrived. The
coincidence window is therefore 400 ms, which corresponds to a coincident false
alarm rate of approximately one 400 ms/(3.4 hr$)^2$, or about one per 12
years. However, recall that the time resolution of the WVD is less than 0.0008
s. Furthermore, recall that Steger's algorithm identifies the position of
ridges in the image (in fact, this feature was implemented in our code
although we did not use it). Thus, at no extra computational cost, one could
set $\tau = \tau_c =$ 10 ms. The TF method can therefore easily achieve a
coincident false alarm rate of $10 \,{\rm ms}/(3.4 \,{\rm hr})^2$, or about
one every $475$ years.

\subsection{False Dismissal Probabilities}
The false dismissal probability depends on the nature and strength of the
signal in the data. We've estimated these probabilities for simulated BBHC
waveforms (Section \ref{subsec:Noise}). We used signals with SNRs (as measured
by optimal filters) in the range 7 to 14. To demonstrate the robustness of
this TF method we selected various coalescence waveforms corresponding to
different binary system masses. We confined ourselves to the (total system)
mass range $45M_\odot$ to $70M_\odot$. In this range the merger phase, for
which the waveform cannot be calculated analytically, sweeps through the
frequency band of maximum sensitivity for initial LIGO detectors, and
dominates the detectable signal\cite{F&H}. These are therefore masses for 
which robust methods will be most useful.

\begin{figure}
\epsffile{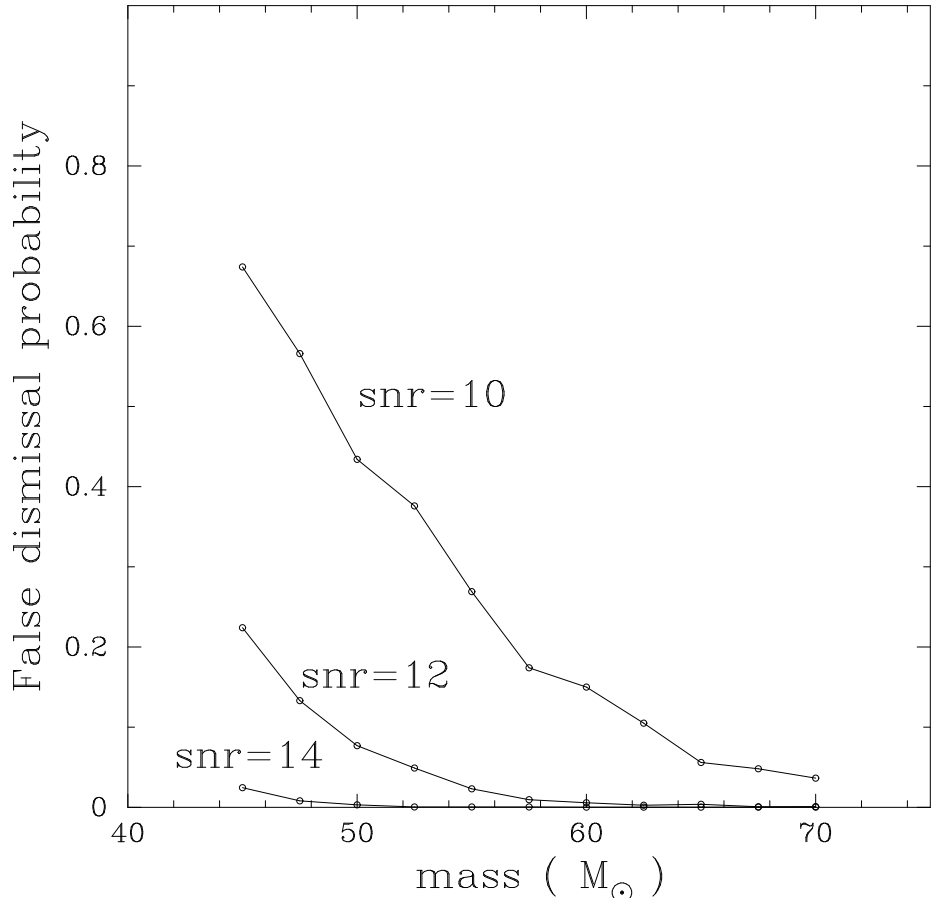}
\caption{False dismissal probability as a function of mass. The three curves
correspond to three different values the optimal filter signal-to-noise ratio.
With the parameters we have chosen, our method tends to work better for higher
mass binaries, where the energy is more localized in the TF map.
\label{fig:dis_vs_mass}}
\end{figure}

Figure \ref{fig:dis_vs_mass} shows false dismissal probabilities as a function
of binary mass at SNRs of 10, 12 and 14. 
At an SNR of 14 the false dismissal rates are acceptable throughout the tested
mass range. At an SNR of 10, however, more than 20\% of signals are missed by
our detection algorithm for system masses
\raisebox{-0.8ex}{$\stackrel{<}{\sim}$} $55 M_\odot$. This is because as the
mass of a system decreases, the proportion of the SNR which is attributable to
the inspiral phase increases, as does the duration of the observable portion
of the inspiral. Thus, the SNR of the signal is spread through a larger region
of the TF map, leading to lower ridges and hence a loss of detectability.  

\begin{figure}
\epsffile{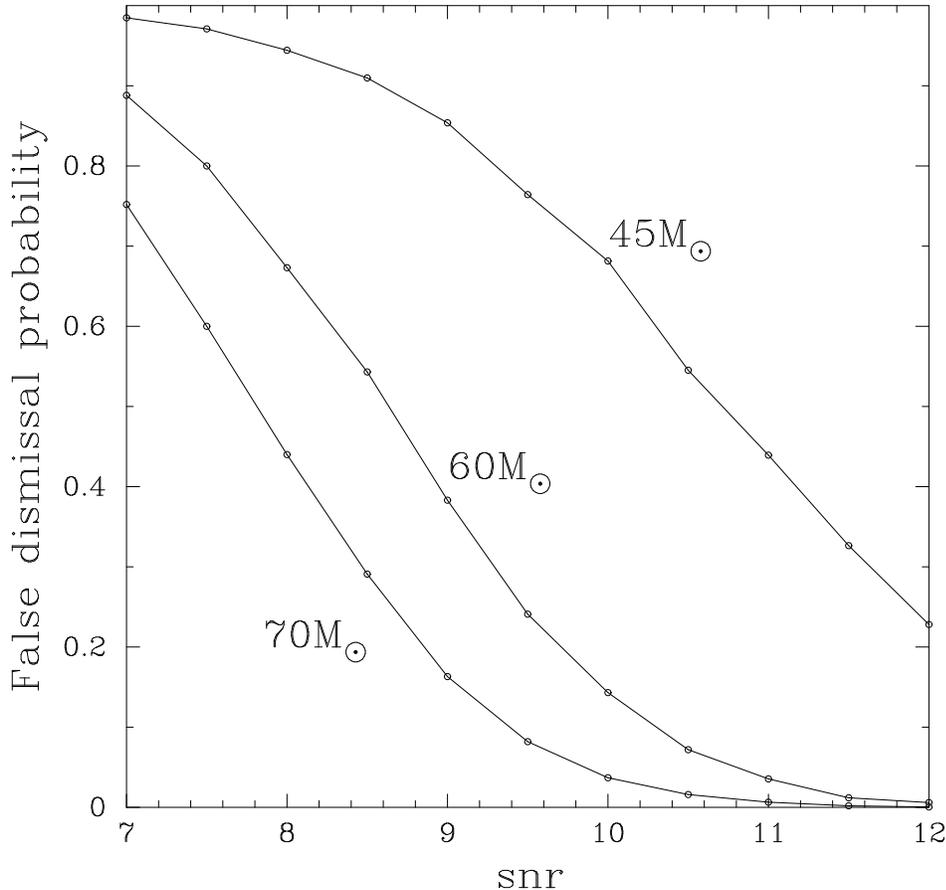}
\caption{False dismissal probability as a function of SNR for three values of
the total mass $M$ of the binary system for the model signal. Even at a modest
SNR of 11, the false dismissal rate is below 50\% for systems with masses
between $70 M_\odot$ and $45 M_\odot$.
\label{fig:dis_vs_SNR}}
\end{figure}

In Figure \ref{fig:dis_vs_SNR} we plot the false dismissal probabilities
versus the SNR for signals from $45M_\odot, 60M_\odot$ and $70M_\odot$ binary
systems. Again, one sees in this Figure that high mass systems are more
readily detected. False dismissal probabilities for every system mass decrease
with signal-to-noise ratio as expected.

\subsection{Computational Efficiency}\label{ss:CompEff}
All our simulations were carried out on a 48 node Beowulf (parallel) computer
comprised 300 MHz Alpha 21164 machines \cite{webpageforbeowulf}. This
computer can analyze 47 segments of data simultaneously (the 48th machine is
used to coordinate the calculation). We found that we could apply our TF
method to approximately $2\times10^4$ data points per second (about twice the
sampling rate of the simulated data). 

For initial LIGO, the frequency band of maximum sensitivity is projected to be
below a few hundred Hz\cite{Abramovicietal}.  Sampling rates of
\raisebox{-.8ex}{$\stackrel{<}{\sim}$} 1024 Hz should therefore be sufficient
to contain all signals detectable with this TF method. Thus, even modest
parallel computing facilities could implement this method in real time for
initial LIGO data. 

\section{Conclusion}\label{sec:conc}
\subsection{Summary}\label{subsec:summ}
Optimal filtering has dominated the interferometric GW data analysis
literature. This is because when one seeks signals of nearly exactly known
form in stationary noise with known spectral density, optimal filters are the
most sensitive linear filters. However, it is clear that there are sources of
gravitational radiation that are not understood well enough to predict their
signal waveforms. For such sources, matched filtering will not be a suitable
detection method. It is therefore necessary to develop methods which can
detect signals from poorly modeled or even unmodeled sources in
interferometric GW data.  

We have demonstrated that time-frequency methods may provide a tool for
detecting such sources. Our method, with the parameters discussed above,
produced an initial LIGO false alarm rate of $\approx (3.4~\mbox{hr})^{-1}$,
which corresponds to coincident false alarm rate in two detectors of $\approx
(475~ \mbox{yr})^{-1}$. Our simulations also show that we reliably detect the
postulated BBHC waveforms for binaries of various masses injected into the
data at a variety of optimal filter signal-to-noise ratios.  For instance, we
detect 80\% of signals with SNRs ranging from $\approx 8.8$ for higher mass
binaries ($70 M_\odot$) to $\approx 12.2$ for lower mass binaries ($45
M_\odot$). These numbers are within a factor of two of those for optimal
filtering of stellar mass binary inspirals, where for two detectors the
detection threshold is usually taken to be an SNR of $\approx 6$ \cite{C&F}.
Furthermore, this method is implementable in real time with a realistic
computing budget.

\subsection{Future Directions}\label{subsec:future}
While this method shows considerable promise, there are still a number of
issues which must be addressed in order to determine how useful TF
methods will be in general:
\begin{description}
\item[Utilization -] One might use TF methods as a filter to identify 
data which should be analyzed for the presence of signals using other
techniques (i.e. in an hierarchical search), for detecting the signals 
themselves, or as a method for characterizing detector noise. This paper only
addresses signal detection.
\item[Choice of Algorithm -] We have presented only one of many TF methods
that could be implemented. We have used the Wigner-Ville distribution and
Steger's ridge detection algorithm, however there are a number of other TFDs
(cf. \cite{GFCM}) and ridge detection algorithms (e.g. Hough
transforms\cite{Hough,DuHa} or the curve and edge detection algorithms cited
in the introduction of \cite{steger98}) which might prove suitable. As a
detection statistic, energy along ridges, total ridge length in a map, or
``bunching'' of ridges might provide better detection statistics than maximum
ridge length. 
\item[Optimization -] We have made heuristic choices for operating values of
the smoothing parameter $\sigma$ and the second derivative thresholds in 
Steger's algorithm. It would be desirable to have a process which, given
a specific type of GW signal, optimizes these parameters.
\item[Implementation -] It is unlikely that a single TF method is sufficiently
robust to detect all types of unmodeled sources, which means that a set of TF
methods should be used. It is also possible that TF methods are optimally
implemented hierarchically. For example, one might have two detection 
statistics; one with an unacceptably high false alarm rate but low
computational cost and another with prohibitive computational cost but a low
false alarm rate. By applying the second statistic only to those TF maps in
which a signal is detected with the first statistic, one could achieve both
computational efficiency and acceptable false alarm rates.
\item[Comparison with Other Techniques -] We do not know how TF methods
compare to other techniques. Signals that are optimal for TF detection produce
regions of high density in the TFD. For our simulated signals, this high
density region spans relatively short time scales ($\approx 0.2$ s). They
might therefore be equally well detected with much simpler time-series
thresholding techniques such as the one described in \cite{ACDH}. We have
investigated this possibility by comparing each datum used in our TF
simulations to a threshold. We chose the threshold to produce the same false
alarm rate as obtained with the TF method. This technique produced a false
dismissal rate of about twice that of the TF method. However,  it is unknown
whether a more sophisticated time-series threshold or a different statistic
such as the excess power statistic \cite{excesspow} would prove superior to
the TF method.  
\end{description}

There is one further issue which pertains to all robust methods. In order to
detect a signal of unspecified form in noise, the noise must be well 
characterized. We have assumed Gaussian noise in this paper. It is uncertain 
at this time to what extent this assumption is valid, however it is known
that noise from the LIGO 40 m prototype contains a significant non-Gaussian 
component. The extent to which this component can be identified and removed
(perhaps with the aid of TF methods themselves) is an issue that will only be
fully resolved once the relevant interferometer data becomes available.

Clearly, there is a much work to be done. Moreover, with interferometric
detectors scheduled to begin taking data in approximately a years time, there
is little time in which to do it. In an effort to expedite the resolution 
of the issues listed above, we have made our TF computer code available as 
part of the latest release of GRASP\cite{GRASP}. 

However, the results of this preliminary investigation are encouraging. With
the resolution of these issues, TF methods promise to be useful tools for the
detection of GW signals in interferometric data. 

\acknowledgements
We would like to thank Bruce Allen for many useful discussions and
suggestions. We also thank Patrick Brady, Eric Chassande-Mottin, 
Jolien Creighton, \'{E}anna Flanagan, Scott Hughes and Alan Wiseman for 
their helpful suggestions, and Carsten Steger for suggesting his 
algorithm and supplying his C code to us. WGA acknowledges the 
financial support of an NSERC Postdoctoral Fellowship.

\appendix
\section{The Binary Black Hole Coalescence Waveform} \label{sec:N&S}
The signal waveforms we used in our false dismissal simulations are based on
the description of binary black hole coalescence (BBHC) signals in Flanagan
and Hughes (FH)\cite{F&H}. They consist of three consecutive components: an
inspiral component, a merger component, and a quasi-normal ringdown component.
The division into these components is somewhat arbitrary. Roughly speaking: 
\begin{itemize}
\item The inspiral component describes the evolution of the binary system up
to the ``innermost stable circular orbit'' (ISCO). This component is well
modeled for BBH systems of $\sim 1 M_\odot$\cite{BIWW}, but there is some
doubt whether it is known well enough for optimal filtering for
binaries of total mass \raisebox{-.8ex}{$\stackrel{>}{\sim}$} 
$20 M_\odot$\cite{BCT}.
\item The merger component describes the evolution of the system from ISCO
until such time as the system can be described as a single perturbed Kerr
black hole. No analytical descriptions of this component exist.
\item The quasi-normal ringdown component describes the evolution of the 
system when it is well described as a single perturbed Kerr black hole. 
This component well modeled for all BBHC's\cite{TeukPress,ChandraDet}.
\end{itemize}
In this Appendix, we model each component separately. The full coalescence
waveform is then simply a combination of the three component waveforms:
\begin{equation}
   \langle h^{\rm BBH}(t) \rangle = \left\{
      \begin{array}[c]{ll} 
         \langle h^{\rm insp}(t) \rangle,&t<0,\\
         A^{\rm merge}(t)~\cos \varphi^{\rm merge}(t),&0 \le t 
            \le t_{\rm merge},\\ 
         \langle h^{\rm qnr}(t) \rangle,&t>t_{\rm merge},
      \end{array} \right.
\label{eq:hBBH}
\end{equation}
An example of such a waveform for a $30 M_\odot$ -- $30 M_\odot$ system is
shown in Figure \ref{fig:BBHwave}. We use units in which $c=G=1$ throughout 
the remainder of this Appendix.

\subsection{Quasi-Normal Ringdown}
The ringdown waveform is given in \cite{F&H} as
\begin{equation}
	h_+^{\rm qnr} - i h_\times^{\rm qnr} = \frac{{\cal A} M}{r}
   ~{}_2S^2_2(\iota,\beta,a) ~e^{-2i\pi f^{\rm qnr} t -t/\tau+i\varphi_0}. 
\label{eq:FH3.15}
\end{equation}
Here, $M$ and $a$ are the mass and angular momentum per unit mass of the black
hole, $r$ is the distance to the hole, ${}_2S^2_2(\iota,\beta,a)$ is a
spin-weighted spheroidal harmonic, $f^{\rm qnr}$ is the characteristic
frequency of the black hole normal mode, $\tau$ is the time constant for the
exponential decay of the mode, $\varphi_0$ is an arbitrary phase factor, and
$\iota$ and $\beta$ are angles that describe the orientation of the black
hole's spin axis with respect to the plane of the detector arms (c.f. Fig. 9.2
of \cite{300y}). This waveform is quite straightforward except for the
presence of the spin-weighted spheroidal harmonic, ${}_2S^2_2(\iota,\beta,a)$. 

This complication of the ${}_2S^2_2(\iota,\beta,a)$ factor is easily
circumvented by considering only the root-mean-squared (RMS) average of it
over the source orientation angles $\iota$ and $\beta$.  When averaged in this
way, spin-weighted spheroidal harmonic obeys the trivial relationship 
\begin{equation}
   \sqrt{\frac{1}{4\pi}\int d\Omega~\left| {}_2S^2_2(\iota,\beta,a) \right|^2}
   =\frac{1}{\sqrt{4\pi}}. \label{eq:FH3.16}
\end{equation}
Since we are not interested in modeling the waveform exactly (indeed, our 
point is to show that we can find even inaccurately modeled signals), let us
for the sake of simplicity substitute the RMS value of $(4\pi)^{-1/2}$ 
for ${}_2S^2_2(\iota,\beta,a)$ in (\ref{eq:FH3.15}) to obtain
\begin{equation}
	\langle h_+^{\rm qnr} \rangle_{(\iota,\beta)} - i \langle h_\times^{\rm qnr} 
   \rangle_{(\iota,\beta)} \equiv \frac{{\cal A} M}{\sqrt{4\pi}r} 
   ~e^{-2i\pi f^{\rm qnr} t -t/\tau+i\varphi_0}. 
   \label{eq:FH3.15aa}
\end{equation}
Note that $\langle h_+^{\rm qnr} \rangle_{(\iota,\beta)}$ and $\langle
h_\times^{\rm qnr} \rangle_{(\iota,\beta)}$ are {\em not} the $(\iota,\beta)$
angle averaged RMS values of $h_+^{\rm qnr}$ and $h_\times^{\rm qnr}$
respectively, but are rather the result of averaging ${}_2S^2_2$ over 
$(\iota,\beta)$.

From (\ref{eq:FH3.15}), the detector response to the quasi-normal ringing of a
black hole is obtained by
\begin{equation}
   h(t)=F_+(\theta,\phi,\psi) h_+(t) + F_\times(\theta,\phi,\psi) h_\times(t),
   \label{eq:THY103}
\end{equation}
where $F_+$ and $F_\times$ are the beam pattern functions
\begin{eqnarray}
   F_+(\theta,\phi,\psi)&=&\frac{1}{2}(1+\cos^2\theta)\cos 2\phi\cos 2\psi
      -\cos \theta \sin 2\phi \sin 2 \psi, \label{eq:THY104a}\\
   F_\times(\theta,\phi,\psi)&=&\frac{1}{2}(1+\cos^2\theta)\cos 2\phi\sin 2\psi
      +\cos \theta \sin 2\phi \cos 2 \psi, \label{eq:THY104b}
\end{eqnarray}
and $\theta$, $\phi$ and $\psi$ are angles which describe the detector
orientation\cite{300y}. One may average over detector orientations to
obtain
\begin{eqnarray}
   \left| h(t) \right|_{(\theta,\phi,\psi)} &=& \sqrt{\frac{1}{4\pi^2}
   \int_0^\pi d\theta \sin \theta \int_0^{2\pi} d\phi \int_0^{\pi}d\psi 
      ~h^2(t)},
   \nonumber \\
   &=& \frac{1}{\sqrt{5}}\sqrt{h_+(t)^2+h_\times(t)^2}. \label{eq:rmsh}
\end{eqnarray}
Substituting (\ref{eq:FH3.15aa}) into (\ref{eq:rmsh}) we get
\begin{equation}
   A^{\rm qnr}(t) \equiv \left|\langle h(t)^{\rm qnr} \rangle_{(\iota,\beta)}
      \right|_{(\theta,\phi,\psi)}=
      \frac{{\cal A} M}{\sqrt{20\pi}r}e^{-t/\tau}. \label{eq:hqnravg}
\end{equation} 
Notice that $A^{\rm qnr}(t)$ is no longer a waveform. The source angle averaging
combines $h_+$ and $h_\times$ to in such a way that only the (exponentially
decaying) amplitude envelope remains. To recover the waveform, we 
multiply $A^{\rm qnr}(t)$ by a cosine at the appropriate frequency,
\begin{eqnarray}
   \langle h(t)^{\rm qnr} \rangle &\equiv& A^{\rm qnr}(t-t_{\rm merge})
      ~\cos(2\pi f^{\rm qnr} (t-t_{\rm merge}) +\varphi_0^{\rm qnr}),
\nonumber \\
   &=&\frac{{\cal A} M}{\sqrt{20\pi}r}e^{-(t-t_{\rm merge})/\tau}
      \cos(2\pi f^{\rm qnr}(t-t_{\rm merge}) +\varphi_0^{\rm qnr}).
\label{eq:hqnr}
\end{eqnarray}
Here, the angle brackets serves to remind us that there have been two averaging
processes used in obtaining this result, and we have offset the time variable
$t$ by $t_{\rm merge}$ (the duration of the merger component) to facilitate
combining it with the other waveform components as in (\ref{eq:hBBH}).

Finally, we wish to express (\ref{eq:hqnr}) in terms of the BBH system mass.
Using Eq. (3.17) of \cite{F&H} and the values $a=0.98$ and ${\cal A}=0.4$ 
quoted in Eq. (3.21) of \cite{F&H}, we have 
\begin{eqnarray}
   f^{\rm qnr} &=& 1320 {\rm Hz} \left(\frac{20 M_\odot}{M}\right),
\label{eq:fh322}\\
   \tau &\approx& \frac{11.63}{\pi f^{\rm qnr}}.
\label{eq:fh3.17}
\end{eqnarray}
Thus, the only free parameters remaining in (\ref{eq:hqnr}) are the system mass
$M$, the time offset $t_{\rm merge}$ and the initial phase of the ringdown
waveform $\varphi_0^{\rm qnr}$. The mass will remain a free parameter in our
simulation, however $t_{\rm merge}$ and the initial phase are determined by 
the merger waveform.

\subsection{Binary Inspiral}
For the binary inspiral
component, $h_+$ and $h_\times$ can be conveniently written in the form 
\cite{300y}
\begin{eqnarray}
   h^{\rm insp}_+(t) &=& -2\left(\frac{\mu}{r}\right)[\pi M f^{\rm insp}(t)]
      ^{2/3} (1+\cos^2 \iota) \cos \varphi^{\rm insp}(t), 
\label{eq:grasp5.6.1} \\
   h^{\rm insp}_\times (t) &=& -4\left(\frac{\mu}{r}\right)[\pi M 
      f^{\rm insp}(t)]^{2/3} \cos \iota ~ \sin \varphi^{\rm insp}(t),
\label{eq:grasp5.6.2}
\end{eqnarray}
where $M=m_1+m_2$ is the total mass of the BBH system, $\mu = (m_1 m_2)/M$ is
the reduced mass of the system, and $f^{\rm insp}(t)$ and $\varphi^{\rm
insp}(t)$ are the binary inspiral orbital phase and frequency. These latter
are given to first post-Newtonian order by\cite{GRASP}
\begin{eqnarray}
   f^{\rm insp}(t)&=&\frac{1}{8 \pi M}\left\{\Theta^{-3/8}+\left(
      \frac{743}{2688} +\frac{11}{32} \eta \right) \Theta^{-5/8}\right\},
\label{eq:grasp5.41} \\
   \varphi^{\rm insp}(t)&=&\varphi_{\rm coal}-\frac{1}{\eta}\left\{\Theta^{5/8}+
      \left(\frac{3715}{8064}+{55}{96}\eta\right)\Theta^{3/8}\right\},
\label{eq:grasp5.42}
\end{eqnarray}
where
\begin{eqnarray}
   \eta &\equiv& \frac{\mu}{M},
\label{eq:grasp5.43a} \\
   \Theta &\equiv& \frac{\eta}{5M}(t_{\rm coal}-t).
\label{eq:grasp5.43}
\end{eqnarray}
$\varphi_{\rm coal}$ and $t_{\rm coal}$ are the phase and time at which
coalescence occurs (i.e. when the frequency becomes infinite). Note that
inspiral waveforms are available to post-Newtonian order 2.5\cite{GRASP}, but
first post-Newtonian order will be sufficient for this crude model.

As with the ringdown waveform, we perform an RMS averaging over the 
source angles in the inspiral waveform. The equivalent averaging in this 
case replaces $(1+\cos^2 \iota)/2$ in $h_+(t)$ and $\cos \iota$ in 
$h_\times(t)$ by their $(\iota,\beta)$ averaged RMS values,
\begin{equation}
   \sqrt{\frac{1}{4\pi}\int d \Omega \left[\frac{1}{4}(1+\cos^2 \iota)^2 + 
      \cos^2 \iota \right]}=\frac{2}{\sqrt{5}},
\label{eq:insaa}
\end{equation}
Thus, 
\begin{eqnarray}
   \langle h^{\rm insp}_+(t)\rangle_{(\iota,\beta)} &=& -8\left(
      \frac{\mu}{\sqrt{5}r} \right)[\pi M f^{\rm insp}(t)]^{2/3} \cos 
      \varphi^{\rm insp}(t), 
\label{eq:grasp5.6.1aa} \\
   \langle h^{\rm insp}_\times (t)\rangle_{(\iota,\beta)} &=& -8\left(
      \frac{\mu}{\sqrt{5}r}\right)[\pi M f^{\rm insp}(t)]^{2/3} \sin 
      \varphi^{\rm insp}(t),
\label{eq:grasp5.6.2aa}
\end{eqnarray}
We also average over the detector orientations. Inserting
(\ref{eq:grasp5.6.1aa}) and (\ref{eq:grasp5.6.2aa}) into (\ref{eq:rmsh}) and
multiplying by $\cos \varphi^{\rm insp}(t)$ we have
\begin{equation}
   \langle h^{\rm insp}(t) \rangle \equiv A^{\rm insp}(t)~\cos 
      \varphi^{\rm insp}(t)= \frac{8\mu}{5r}[\pi M f^{\rm insp}(t)]^{2/3} 
      \cos \varphi^{\rm insp}(t).
\label{eq:hinsp}
\end{equation}

Again, we wish to express the waveform in terms of the system mass $M$. It
therefore remains to fix the reduced mass $\mu$, the coalescence time $t_{\rm
coal}$, and the coalescence phase $\varphi_{\rm coal}$. We restrict our
attention to equal mass binaries, so that the reduced mass is therefore
$\mu=M/4$. We fix $t_{\rm coal}$ so that the binary system attains the ISCO
frequency, $f_{\rm ISCO}$, at time $t=0$, i.e. so that $f^{\rm insp}(0)=
f_{\rm ISCO}$.  Following \cite{F&H}, we use the Kidder-Will-Wiseman\cite{KWW}
value of $f_{\rm ISCO} = (20 M_\odot/M) \times 205 {\rm Hz}$. We also fix
$\varphi_{\rm coal}$ so that the phase of the binary system vanishes at $t=0$,
i.e. so that $\varphi^{\rm insp}(0)=0$. Note that this is {\em not} the same
as choosing $t_{\rm coal}=0$ and $\varphi_{\rm coal}=0$, since coalescence
occurs after the ISCO.  This completes the specification (apart from the free
mass parameter) of the inspiral component of the BBHC waveform.

\subsection{Merger}
The remaining task is to model the merger waveform. While no analytic model
for the merger exists, Flanagan and Hughes\cite{F&H} make educated estimates
of some of its properties. They assume that the merger signal contains only
frequencies between the ISCO frequency, $f_{\rm ISCO}$, and the quasi-normal
ringing frequency, $f^{\rm qnr}$. They also estimate that the energy carried
by the merger component of the wave is approximately $3$ times the energy
carried by the quasi-normal ringdown component. Finally, they estimate the
duration of the merger to be $t_{\rm merge}\sim 50M$. We use these criteria,
along with the requirement that $h$ and $\partial_t h$ be continuous, to guide
us in making a mock merger waveform. While the resulting waveform won't be
that of a real BBH merger, it should resemble it enough to determine whether
our TF method can detect BBH waveforms for intermediate-mass systems. 

The first step in our construction of a merger waveform estimate is to assume
a form of
\begin{equation}
   \langle h^{\rm merge}(t) \rangle = A^{\rm merge}(t) \cos 
      \varphi^{\rm merge}(t).
\label{eq:mergeansatz}
\end{equation}
Continuity of the $f_{\rm merge}$ and $\partial_t f_{\rm merge}$ with both
the inspiral waveform that proceeds it and the quasi-normal ringdown waveform
that follows constitutes four conditions on $\varphi^{\rm merge}(t)$:
\begin{eqnarray}
   \frac{\di \varphi^{\rm merge}}{\di t}(t=0)&=&f_{\rm ISCO},
\label{eq:phicond1}\\
   \left(\frac{\di^2 \varphi^{\rm merge}}{\di t^2}\right)(t=0)&=&
      \left(\frac{\di^2 \varphi^{\rm insp}}{\di t^2}\right)(t=0),
\label{eq:phicond2}\\
   \frac{\di \varphi^{\rm merge}}{\di t}(t=t_{\rm merge})&=&
   f^{\rm qnr}, 
\label{eq:phicond3}\\
   \left(\frac{\di^2 \varphi^{\rm merge}}{\di t^2}\right)
      (t=t_{\rm merge})&=& 0,
\label{eq:phicond4}
\end{eqnarray}
where the last equation follows from the fact that the $f^{\rm qnr}$ is
constant. To satisfy these four conditions will require a frequency model with
four free parameter. We use the simplest such model; the merger frequency a
cubic function of time. The phase is therefore a quartic of the form
\begin{equation}
   \phi^{\rm merge} (t) = f_{\rm ISCO} t + 
      \left(\frac{\di^2 \varphi^{\rm insp}}{\di t^2}(0)\right) t^2 +
      \varphi^{\rm merge}_3 t^3 + \varphi^{\rm merge}_4 t^4,
\label{eq:phimerge}
\end{equation}
where $\varphi^{\rm merge}_3 $ and $\varphi^{\rm merge}_4$ can be obtained by
solving (\ref{eq:phicond3}) and (\ref{eq:phicond4}) and we have set
$\varphi^{\rm merge} (0)=0$. Once the merger phase polynomial has been
determined, it is a simple matter to find the phase constant for the
quasi-normal ringdown,
\begin{equation}
   \varphi^{\rm qnr}_0 = \varphi^{\rm merge} (t=t_{\rm merge}),
\label{eq:phiqnr0}
\end{equation}
which completes the specification of the quasi-normal ringdown component of
the waveform.

To determine the merger amplitude function, we impose similar continuity 
conditions 
\begin{eqnarray}
   A^{\rm merge}(t=0)&=&A^{\rm insp}(t=0) 
\label{eq:Acond1}\\
   \left(\frac{\di A^{\rm merge}}{\di t}\right)(t=0)&=&
      \left(\frac{\di}{\di t} A^{\rm insp} \right)(t=0),
\label{eq:Acond2}\\
   A^{\rm merge}(t=t_{\rm merge})&=& A^{\rm qnr} (t=t_{\rm merge}),
\label{eq:Acond3}\\
   \left(\frac{\di A^{\rm merge}}{\di t}\right)(t=t_{\rm merge})&=&
      \left(\frac{\di}{\di t} A^{\rm qnr}\right) (t=t_{\rm merge}),
\label{eq:Acond4}
\end{eqnarray}
We again need a fitting function with at least four parameters.  However,
there is a further constraint to impose; recall that \cite{F&H} estimates the
energy of the merger to be three times the energy of the quasi-normal
ringdown, i.e.
\begin{equation}
   \int_0^{t_{\rm merge}} \left(\frac{\di}{\di t}  
      \langle h^{\rm merge}(t) \rangle \right)^2 dt = 
      3 \int_{t_{\rm merge}}^\infty \left(\frac{\di}{\di t}  
      \langle h^{\rm qnr} (t) \rangle \right)^2 dt.
\label{eq:mergenergy}
\end{equation}
This fifth condition on $A^{\rm merge}(t)$ requires a fifth parameter, and 
we therefore model the merger amplitude with a quartic,
\begin{equation}
   A^{\rm merge}(t)= A^{\rm insp}(0) + \left(\frac{\di}{\di t} A^{\rm insp} 
      \right)(0)~t+A^{\rm merge}_2 t^2+A^{\rm merge}_3 t^3
      +A^{\rm merge}_4 t^4,
\label{eq:Amerge} 
\end{equation}
where the coefficients $A^{\rm merge}_{2,3,4}$ are determined by
(\ref{eq:Acond3}), (\ref{eq:Acond4}) and (\ref{eq:mergenergy}). This completes
the specification of the merger waveform.


\end{document}